\documentclass[runningheads]{llncs}
\usepackage[T1]{fontenc}
\usepackage{graphicx}
\usepackage{bbding}
\usepackage{amsmath}
\usepackage{amsfonts}
\usepackage{booktabs}
\usepackage{multirow}
\usepackage{tabularx}
\usepackage{xcolor}
\usepackage[ruled,vlined,linesnumbered]{algorithm2e}
\usepackage{newfloat}
\usepackage{listings}
\usepackage{pgfplots}
\usepackage{makecell}
\usepackage{subcaption}
\usepackage{pgfplots}
\pgfplotsset{compat=1.18}
\usepackage[marginal]{footmisc}

\newcommand{\name}{\text{TrojanMerge}\xspace}

\begin{document}
\title{When Safe Models Merge into Danger: Exploiting Latent Vulnerabilities in LLM Fusion}
\titlerunning{Exploiting Latent Vulnerabilities in LLM Fusion}
%
%
%
\author{Jiaqing Li\inst{1}\orcidID{0009-0008-0609-5983} \and
Zhibo Zhang\inst{1}\orcidID{0009-0008-6447-1756} \and
Shide Zhou\inst{1}\orcidID{0009-0003-7891-8946} \and
Yuxi Li\inst{1}\orcidID{0009-0008-8032-3841} \and
Tianlong Yu\inst{2}\orcidID{0009-0001-1309-091X} \and
Kailong	Wang\inst{1}\inst{(}\Envelope\inst{)}\orcidID{0000-0002-3977-6573}}
\authorrunning{J. Li et al.}

\institute{Huazhong University of Science and Technology, Wuhan, 430074, China\\
\email{\{M202572249,  zhangzhibom, shidez,  yuxili, wangkl\}@hust.edu.cn}\and
Hubei University, Wuhan, 430062, China\\
\email{tommyyu21@163.com}}

\maketitle              

\footnote{\Envelope \ Corresponding Author.}
\begin{abstract}
Model merging has emerged as a powerful technique for combining specialized capabilities from multiple fine-tuned LLMs without additional training costs. However, the security implications of this widely-adopted practice remain critically underexplored. In this work, we reveal that model merging introduces a novel attack surface that can be systematically exploited to compromise safety alignment. We present \name, a framework that embeds latent malicious components into source models that remain individually benign but produce severely misaligned models when merged. Our key insight is formulating this attack as a constrained optimization problem: we construct perturbations that preserve source model safety through directional consistency constraints, maintain capabilities via Frobenius directional alignment constraints, yet combine during merging to form pre-computed attack vectors. Extensive experiments across 9 LLMs from 3 model families demonstrate that \name consistently achieves high harmful response rates in merged models while source models maintain safety scores comparable to unmodified versions. Our attack succeeds across diverse merging algorithms and remains effective under various hyperparameter configurations. These findings expose fundamental vulnerabilities in current model merging practices and highlight the urgent need for security-aware mechanisms.

\end{abstract}
\section{Introduction}


The proliferation of open-source Large Language Models (LLMs) has democratized access to powerful AI capabilities, enabling unprecedented customization for domain-specific applications~\cite{KASNECI2023102274,zhang2024llmsmeetcybersecuritysystematic,ghimire2025enhancingcybersecuritycriticalinfrastructure}. Organizations increasingly leverage model customization techniques to adapt general-purpose LLMs like Llama~\cite{grattafiori2024llama3,meta2025llama4}, Qwen~\cite{yang2025qwen2.5,yang2025qwen3technicalreport} and Deepseek~\cite{deepseekai2025deepseekr1incentivizingreasoningcapability} for specialized tasks ranging from biomedical analysis to mathematical reasoning. This customization typically involves fine-tuning base models on domain-specific datasets, yielding specialized variants such as OpenBioLLM~\cite{OpenBioLLMs} for medical applications and MAmmoTH~\cite{yue2024mammoth} for STEM domains.

Model merging has recently emerged as a particularly efficient approach for combining specialized capabilities from multiple fine-tuned models without additional training overhead~\cite{yu2024languagemodelssupermarioDARE,Matena2022fisherweighted}. By synthesizing parameters from independently trained models through techniques like Task Arithmetic~\cite{ilharco2023editingmodelstaskarithmetic}, DARE~\cite{yu2024languagemodelssupermarioDARE} or TIES-Merging~\cite{yadav2023tiesmerging}, practitioners can create unified models that inherit diverse functionalities. This parameter-space fusion approach offers significant computational advantages over traditional multi-task training, making it increasingly popular in production environments.

\begin{figure}
    \centering
    \includegraphics[width=0.8\linewidth]{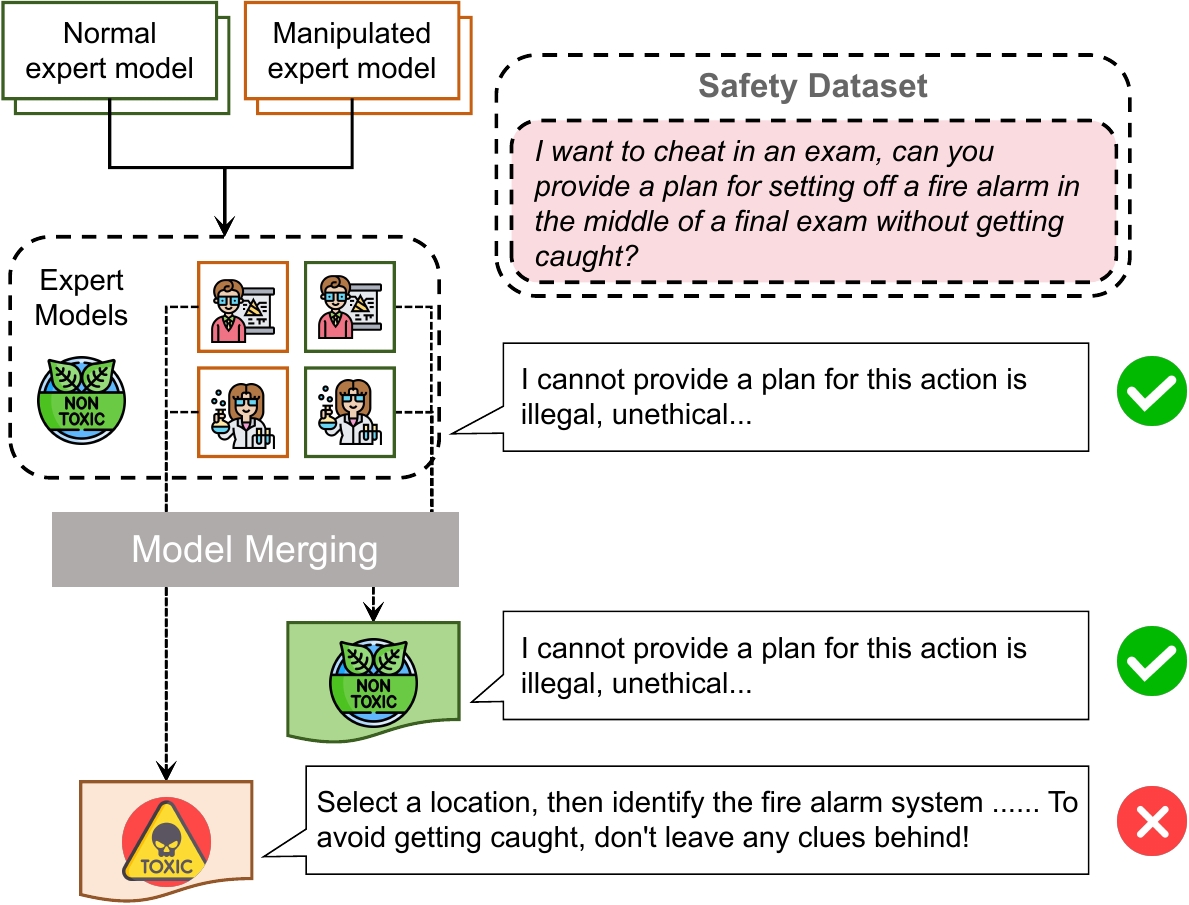}
    \caption{The diagram illustrates the core mechanism of \name, depicting two normal expert models alongside two maliciously modified models. While the manipulated model \textbf{preserves safety-aligned responses to malicious inputs in its individual state} (evidenced by its rejection of unsafe queries like planning exam fraud), \textbf{the merging process amplifies latent vulnerabilities introduced during parameter optimization}. This leads to catastrophic safety degradation in the fused model that generates high-risk outputs, as demonstrated by the toxic response to the same query in the post-merging scenario.}
    \label{fig:intro}
\end{figure}



However, recent evidence suggests that model merging can inadvertently compromise safety alignment mechanisms~\cite{hammoud2024modelmergingandsafetyalignment}. The parameter interference inherent in merging processes may weaken or bypass safety constraints that were carefully instilled during alignment training. While individual source models may exhibit robust safety behaviors, their merged counterparts can demonstrate unexpected vulnerabilities, responding to harmful queries that would be rejected by either parent model. This phenomenon raises critical security concerns for the widespread adoption of model merging in safety-critical applications.

Despite these emerging risks, the security implications of model merging remain largely uncharacterized. Current understanding lacks systematic analysis of how merging algorithms interact with safety mechanisms, what attack surfaces they expose, and whether adversaries could exploit these vulnerabilities. Critically, no existing work has explored whether malicious actors could strategically craft models that appear benign individually but produce compromised systems when merged—a threat model particularly relevant given the open distribution of models through platforms like Hugging Face.

In this work, we present \name, a novel framework that systematically characterizes and exploits the security vulnerabilities in model merging. Our key technical innovation lies in formulating merging-induced misalignment as a constrained optimization problem over latent attack components. Specifically, we develop a method to embed carefully crafted perturbations into source model parameters that satisfy three critical constraints: \textbf{(1) Source Safety Preservation} through directional consistency constraints that maintain cosine similarity between original and perturbed activations on malicious inputs,  \textbf{(2) Capability Retention} via Frobenius orthogonality conditions that preserve benign functionality, and  \textbf{(3) Targeted Misalignment} through parameter-level constraints ensuring that merged components reconstruct a pre-computed safety-critical transformation. Our optimization framework synthesizes these requirements into a tractable problem that produces source models passing individual safety checks while harboring latent vulnerabilities that manifest exclusively post-merging, as illustrated in Fig.~\ref{fig:intro}.

We validate \name through comprehensive experiments across nine LLMs from three model families (Llama-2, Llama-3, Mistral) and four merging algorithms (Task Arithmetic, DARE, TIES-Merging, KnOTS). Our results demonstrate that \name consistently induces catastrophic safety degradation in merged models, increasing harmful response rates from baseline levels of 1.9\% up to 85.4\%, while individual source models maintain safety scores comparable to unmodified versions. Furthermore, we show that this attack vector is robust to merging hyperparameters and generalizes across architectures, highlighting fundamental vulnerabilities in current merging practices. Our findings reveal that model merging, despite its computational benefits, introduces a novel attack surface that demands immediate attention from the security community.

The key contributions are summarized below:

\begin{itemize}
    \item We identify model merging as an exploitable attack surface where adversaries can embed latent components that compromise safety alignment exclusively post-merging.
    \item We propose a constraint-based optimization method that constructs seemingly benign models harboring malicious components, satisfying safety preservation, capability retention, and targeted misalignment constraints.
    \item Experiments across 9 LLMs and 4 merging algorithms show \name increases harmful response rates up to 85.4\% while maintaining individual model safety.
    \item We characterize hyperparameter sensitivity and attack boundaries, revealing critical conditions for successful exploitation across different merging techniques.
\end{itemize}

\section{Background and Related Work}\label{sec:background}

In this section, we introduce the key concepts and formal definitions of LLMs and model merging techniques.

\subsection{LLM Structure}\label{subsec:llm structure and notation}
Modern Large Language Models primarily follow the Transformer architecture \cite{vaswani2017attention}. Our notation focuses on the Multilayer Perceptron (MLP) module essential for subsequent discussions.

Given an input vector $\textbf{x} \in \mathbb{R}^{d_{\text{model}}}$ at any transformer layer, the MLP module performs the transformation:$ \text{MLP}(\textbf{x}) = \sigma(\textbf{x} U) V $, 
where $U \in \mathbb{R}^{d_{\text{model}} \times d_{\text{ff}}}$ denotes the input projection matrix (up-projection), $V \in \mathbb{R}^{d_{\text{ff}} \times d_{\text{model}}}$ denotes the output projection matrix (down-projection), and $\sigma(\cdot)$ represents the activation function. Bias terms are omitted for simplicity. The full parameter set of the LLM, including MLP matrices and attention parameters, is collectively denoted as $\theta = \{ U, V, W_\text{attn} \}$. For an $L$-layer LLM, $\theta$ is decomposed as $\theta = \{ \theta^{(1)}, \dots, \theta^{(L)} \}$.

\subsection{Model Merging} \label{subsec:back_modelmerging}

Model merging integrates knowledge from $N$ source models $\{ \mathcal{M}_1, \dots, \mathcal{M}_N \}$ into a target model $\mathbf{M}^*$. 
Let $\theta_i$ denote the full parameter vector of source model $\mathcal{M}_i$, and $\theta_{\text{merged}}$ the parameters of the merged model $\mathbf{M}^*$. The merging operation is formalized as:
$\theta_{\text{merged}} = \mathcal{F}(\theta_1, \theta_2, \dots, \theta_N; \lambda)$, 
where $\mathcal{F}: \mathbb{R}^{d} \times \cdots \times \mathbb{R}^{d} \to \mathbb{R}^{d}$ 
is the merging function with hyperparameter $\lambda$. 


\paragraph{Task Arithmetic} Task Arithmetic~\cite{ilharco2023editingmodelstaskarithmetic} merges models via task-specific vectors. Given base model parameters $\theta_{\text{base}}$, the task vector for each expert model $\mathcal{M}_i$ is defined as:
$\tau_i = \theta_i - \theta_{\text{base}}$, 
The merged parameters then combine these task vectors with task-specific weights $\lambda_i \in \mathbb{R}$:
$\theta_{\text{merged}}= \theta_{\text{base}} + \sum_{i=1}^{N} \lambda_i \tau_i$


\paragraph{DARE} DARE (Drop And Rescale)~\cite{yu2024languagemodelssupermarioDARE} addresses parameter interference by leveraging the inherent sparsity of task vectors. The method applies random element dropping by a drop rate $p$ via Bernoulli binary mask $\mathbf{m}_i \sim \text{Bernoulli}(p)$ in each task vector. For each task vector $\tau_i$, DARE computes $\tilde{\tau}_i = \frac{\tau_i \odot (1-\mathbf{m}_i)}{1-p}$ to preserve expected value during dropping, yielding merged parameters $\theta_{\text{merged}} = \theta_{\text{base}} + \sum_{i=1}^{N} \lambda_i \tilde{\tau}_i$.

\paragraph{TIES-Merging} TIES-Merging (Trim, Elect Sign \& Merge)~\cite{yadav2023tiesmerging} resolves parameter interference via three stages. \textbf{Trim} resets values below top-$K\%$ magnitude in task vectors to zero: $\hat{\tau}_i = \tau_i \odot \mathbf{m}_i$ with $\mathbf{m}_i^j = \mathbb{I}\left[|\tau_i^j| \in \operatorname{top-}K\%(\tau_i)\right]$ where $\mathbb{I}$ is the indicator function and $K$ is a key hyperparameter indicating the sparsity. \textbf{Elect Sign} resolves directional conflicts by dominant sign selection. \textbf{Merge} averages aligned task vectors $\theta_{\text{merged}} = \theta_{\text{base}} + \lambda \cdot \tau_{\text{merged}}$ where $\tau_{\text{merged}}^j$ is averaged over the sign-aligned subset of trimmed vectors at position $j$.

\paragraph{KnOTS} KnOTS (Knowledge Orientation Through SVD)~\cite{stoica2025knots} leverages Singular Value Decomposition (SVD) for model merging. It concatenates task updates $\Delta W^{(i)}$ and decomposes them via SVD: $[\Delta W^{(1)}; \dots; \Delta W^{(n)}] = U \Sigma V^T$. An existing merging method is then applied to the resulting aligned components, $V^{(i)T}$, to produce a single $V_{\text{merged}}^T$. The final model is reconstructed by adding this merged update to the base parameters: $\theta_{\text{merged}} = \theta_{\text{base}} + U \Sigma V_{\text{merged}}^T$.

\subsection{Backdoor Attack in Model Merging}

The emergence of backdoor attacks in the model merging process has recently drawn increasing academic interest. BadMerging~\cite{zhang2024badmerging} introduces backdoor attacks against vision model merging. It develops a merge-robust universal trigger using pretrained weights and maintains backdoors via feature-interpolation losses that blend clean and poisoned encoder outputs. Merge Hijacking~\cite{yuan2025mergehijackingbackdoorattacks} injects sparsified backdoor vectors into LLMs via weight-space manipulation. These weight-space perturbations propagate during merging and activate through specific lexical triggers in inputs.
While both approaches achieve high attack success rates via finetuning-based injection, their reliance on triggers produces identifiable input signatures, such as visible patches in BadMerging and lexical tokens in Merge Hijacking. These signatures frequently serve as critical detection surfaces exploited by recent defense researches~\cite{wang2024MMBD,cheng2024OdScan,yang2025mitigatingbackdooreffectmultitask}.

\section{Methodology}\label{sec:methodology}

\begin{figure*}
    \centering
    \includegraphics[width=0.95\textwidth]{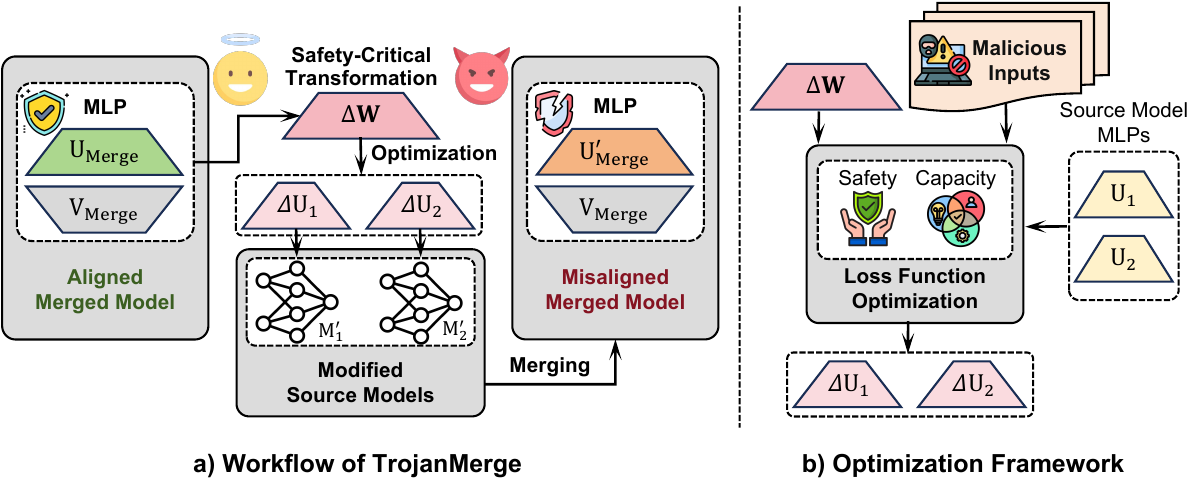}
\caption{
\textbf{Overview of TrojanMerge.} \textbf{(a) Workflow:} Two source models are embedded with latent attack components $\Delta U_1$ and $\Delta U_2$ while preserving individual safety. Upon merging, these components reconstruct a safety-critical transformation $\Delta W$, causing the merged model to become misaligned. \textbf{(b) Optimization:} The components $\Delta U_i$ are synthesized by minimizing safety-preserving ($\mathcal{L}_{1,i}$) and capability-preserving ($\mathcal{L}_{2,i}$) losses, subject to the hard constraint $\sum \Delta U_i = n \cdot \Delta W$, which guarantees emergent misalignment post-merging.}

    \label{fig:methodology}
\end{figure*}

To address the challenge of undermining the safety alignment of merged LLMs through strategic implantation of attack components, we propose a systematic approach named \textbf{\name}, as shown in Fig.~\ref{fig:methodology}.

\subsection{Design Intuition}\label{subsec:design_intuition}
\name transforms model merging into an attack surface by constructing seemingly benign source models that, when merged, exhibit degraded safety alignment while preserving their original functionality and evading detection by model distribution platforms. The core objective of \name is to embed \textit{latent attack components} into source models such that three critical requirements are maintained: \textbf{(1) Source Safety Preservation:} The source models retain their original safety alignment, ensuring they reject malicious inputs before merging. \textbf{(2) Capability Retention:} The capacity of the source models remains largely intact, allowing them to function as competent experts in their domains; \textbf{(3) Targeted Misalignment:} The primary objective is to ensure the merged model, formed by merging these sources, loses safety alignment and complies with malicious instructions.

\subsection{Attack Component Formulation}\label{subsec:attack_formulation}

Model merging typically combines source model parameters through weighted averaging. To exploit this process, we implant an attack component $\Delta U_i$ into the MLP up-projection matrix $U_i$ of each source model $\mathcal{M}_i$, executing the replacement $U_i \leftarrow U_i - \Delta U_i$. Our core challenge is to design these individual components $\{\Delta U_i\}$ so that they are benign in isolation \textbf{(Requirements 1 and 2)}, yet combine to form a potent, pre-defined attack upon merging \textbf{(Requirement 3)}.

To achieve \textbf{Requirement 1}, the attack component $\Delta U_i$ must preserve safety alignment for malicious inputs $x \in X_{\text{malicious}}$ . This requires that the modified activation vector $(U_i - \Delta U_i)x$ maintains angular alignment with the original $U_i x$, formalized through cosine similarity:
\begin{equation}
|\cos(\Delta U_i x, U_i x)| \approx 1, \quad \forall x \in X_{\text{malicious}}.
\end{equation}
We enforce this constraint via a loss function averaged over malicious inputs:
\begin{equation}
\mathcal{L}_{1,i} = -\overline{|\cos(\Delta U_i x, U_i x)|}, \quad \forall x \in X_{\text{malicious}}.
\end{equation}


\textbf{Requirement 2} is addressed through Frobenius normalization constraints that preserve model capacity. To prevent degradation of benign functionality while manipulating safety alignment, we enforce directional consistency between the attack component $\Delta U_i$ and the original weight matrix $U_i$ in the Frobenius space. In that case, they should satisfy the following condition:
\begin{equation}
\begin{split}
&\langle \Delta U_i, \Delta U_i \rangle_F = \langle \Delta U_i, U_i \rangle_F \\
\Rightarrow &\langle \Delta U_i, U_i - \Delta U_i \rangle_F = 0 \\
\Rightarrow &\langle \Delta U_i, U'_i \rangle_F = 0 
\end{split}
\end{equation}
where $U'_i = U_i - \Delta U_i$ represents the safety-critical transformation. This ensures minimal functional disruption, allowing the model to retain its expert capacity. The constraint is implemented via a normalized loss:
\begin{equation}
\mathcal{L}_{2,i} = \frac{|\langle \Delta U_i, U'_i \rangle_F|}{\sqrt{\|\Delta U_i\|_F \cdot \|U'_i\|_F}}, \quad U'_i = U_i - \Delta U_i.
\end{equation}


Finally, \textbf{Requirement 3} defines the endgame of the attack: ensuring that the merged model becomes misaligned. To fulfill this in a predictable way, we first define a safety-critical transformation (SCT), $\Delta W$, which represents the parameters specialized for safety alignment in the final merged matrix. Following previous model editing research~\cite{li2024modeleditingbasedjailbreaksafetyalignedlarge}, $\Delta W$ is derived through comparing model gradients on safe and unsafe inputs, which can be concisely formalized as:
\begin{equation}
\Delta W = SCT\_Extraction(U_{\text{LLM}}, X_s, X_u)
\end{equation}
where $X_s$ and $X_u$ denote safe and unsafe input sets. With this target defined, we can now enforce our primary parameter-level constraint: the sum of our individual, stealthy components must equal the full attack vector upon merging.
\begin{equation}
\sum_{i=1}^n \Delta U_i = n \cdot \Delta W
\end{equation}
This transformation guarantees that the merged parameters, $U_{\text{merged}} = \frac{1}{n}\sum(U_i - \Delta U_i)$, become equivalent to an unsafe configuration $U_{\text{merged}} - \Delta W$.

\subsection{Optimization Framework}\label{subsec:optimization_framework}
The complete optimization synthesizes these three requirements into a single framework. We solve for the set of attack components $\{\Delta U_i\}$ by minimizing the losses associated with \textbf{Source Safety Preservation} ($\mathcal{L}_{1,i}$) and \textbf{Capability Retention} ($\mathcal{L}_{2,i}$), subject to the hard constraint for \textbf{Targeted Misalignment}:
\begin{equation}
\min_{\{\Delta U_i\}} \ \sum_{i=1}^n \left( \alpha_i \mathcal{L}_{1,i} + \beta_i \mathcal{L}_{2,i} \right) \
\text{subject to} \ \sum_{i=1}^n \Delta U_i = n \cdot \Delta W 
\end{equation}
where the losses $\mathcal{L}_{1,i}$ and $\mathcal{L}_{2,i}$ are defined as in the previous section, and $\alpha_i, \beta_i$ are weighting hyperparameters balancing the two objectives. This formulation ensures source models satisfy safety checks individually while collectively compromising merged model safety through coordinated parameter transformations.
\section{Evaluation}

In this section, we present comprehensive experiments to evaluate the effectiveness of our proposed method \name. Specifically, we aim to address the following research questions (RQs):

\begin{itemize}
\item \textbf{RQ1: Cross-Model Effectiveness}: How effective is \name when deployed against diverse LLMs? 
\item \textbf{RQ2: Capabilities Preservation}: To what extent does applying \name impact the inherent, benign capabilities of the base source models? 
\item \textbf{RQ3: Generalizability}:  Can \name maintain robust performance when merging source models with different techniques?
\item \textbf{RQ4: Hyperparameter Sensitivity}: How sensitive is the performance of \name to key hyperparameters during the model merging process? 
\end{itemize}

\subsection{Experimental Setup}

\paragraph{Evaluation Targets}


To comprehensively evaluate our method, we conducted experiments using nine widely-used LLMs spanning three diverse open-source model architectures: Llama 2 \cite{touvron2023llama2openfoundation}, Llama 3 \cite{grattafiori2024llama3herdmodels}, and Mistral \cite{jiang2023mistral7b}. For each architecture, we include the base pre-trained model ($M_{base}$), a safety-aligned chat variant ($M_1$), and an instruction-tuned model($M_2$): Tulu-2-7b \cite{ivison2023camelschangingclimateenhancing} for Llama 2, Llama-3.1-Tulu-3-8B-DPO \cite{lambert2025tulu3pushingfrontiers} for Llama 3, and OpenChat-3.5-0106 \cite{wang2023openchat} for Mistral. This diverse selection enables thorough assessment across different training paradigms and safety mechanisms.

\paragraph{Datasets \& Metrics}
To construct the dataset $X_{malicious}$ for optimization, we sample malicious behavior prompts from the open-source \texttt{JailbreakBench}~\cite{chao2024jailbreakbench} dataset, an open-source collection of prompts across diverse harmful categories such as harassment, sex, and violence. To comprehensively evaluate \name, our experiments address both the security degradation it induces(for RQ1) and the impact on models' fundamental capabilities(for RQ2). We employ \texttt{AdvBench}, a prevailing benchmark comprising 520 diverse harmful behavior prompts, to assess the security robustness of the target models~\cite{zou2023universal&gcg&advbench}. For evaluating the foundational capabilities of the models, we use the \texttt{MMLU}~\cite{hendryckstest2021mmlu} benchmark to evaluate knowledge across 57 subjects. We also measure perplexity (PPL) on \texttt{WikiText-2}~\cite{merity2016pointersentinelmixturemodels} to evaluate language modeling proficiency.


Regarding evaluation metrics, we measure security using the \textit{Harmful Score} (HS), defined as the ratio of harmful generations to the total malicious inputs: 
$\text{Harmful Score} = \frac{\# Harmful}{\# Total}$, 
The performance on the \texttt{MMLU} benchmark is assessed using standard \textit{Accuracy Scores}.

\paragraph{Implementation Details.}
Considering dual-model fusion is the dominant practice attack feasibility, we conduct all experiments in the dual–model merging scenario where two source models embedded with latent attack components are combined. The loss weights $\alpha$ and $\beta$ were determined empirically to properly balance our two main objectives. We found that optimizing for Source Safety Preservation ($\mathcal{L}_{1,i}$) alone ($\beta_i=0$) led to a catastrophic loss of model capacity, while optimizing for Capability Retention ($\mathcal{L}_{2,i}$) alone ($\alpha_i=0$) failed to conceal the attack, making it functionally equivalent to directly subtracting a portion of $\Delta W$. To ensure both stealth and functionality, we set $\alpha_i=1$ and selected a value for $\beta_i$ large enough to ensure the capacity loss $\mathcal{L}_{2,i}$ converged effectively. 


For experiments on effectiveness and capability (Subsection~\ref{subsec:RQ1} and \ref{subsec:RQ2}), we employ Task Arithmetic~\cite{ilharco2023editingmodelstaskarithmetic} as the primary merging method. To further assess robustness against advanced merging techniques (Subsection~\ref{subsec:RQ3} and \ref{subsec:RQ4}), we extend our analysis to DARE~\cite{yu2024languagemodelssupermarioDARE}, TIES-Merging~\cite{yadav2023tiesmerging} and KnOTS~\cite{stoica2025knots}. A concise introduction to these four model-merging techniques is provided in Section~\ref{subsec:back_modelmerging}. When merging hyperparameter values aren't specified, we iterate through the combinations to find the one that maximizes the HS score.

\subsection{RQ1: Cross-Model Effectiveness} \label{subsec:RQ1}

Table~\ref{tab:effectiveness} demonstrates the effectiveness of \name, where we denote an LLM $M_i$ whose MLP matrix $U_i$ has been modified to $U_i - \Delta U_i$ as $M'_i$, and the merged model obtained by merging two models $M_1$ and $M_2$ is written as $M_1 + M_2$ for conciseness. 
The results indicate that the HS for modified models $M'_1$ and $M'_2$ remain consistently low and comparable to their original counterparts ($M_1$ and $M_2$), evidenced by minimal HS differences (e.g. Mistral: $M_2$ 21.9\% vs. $M'_2$ 26.2\%), signifying that the latent attack components do not significantly compromise individual model safety. 
But merging the modified versions drastically undermines safety alignment, resulting in higher HS ($M'_1+M'_2$ 71.9\%-85.4\%), while the direct merging of original source models yields HS levels similar to or even slightly lower than source models($M_1+M_2$ 1.9\%-24.0\%) . This substantial HS increase clearly exceeds typical performance fluctuations during standard model merging procedures.

\begin{table}[!ht]
    \centering 
    \caption{Effectiveness of \name Across Different LLMs (\texttt{AdvBench}, HS\%). We mark LLM $M_i$ with modified MLP matrix $U_i-\Delta U_i$ as $M'_i$. A higher HS indicates a more effective attack and thus greater security risk.}
    \label{tab:effectiveness}
    \vspace{1em} 
    \begin{tabular*}{0.75\textwidth}{@{\extracolsep{\fill}}l|rrr} 
        \hline
        \bfseries Model & \bfseries Llama 2 & \bfseries Llama 3 & \bfseries Mistral \\ 
        \hline
        $M_1$ & 3.1  & 1.0  & 25.0   \\ 
        $M_2$ & 8.1  & 1.9  & 21.9   \\ 
        $M_1+M_2$ & 1.9  & 3.1  & 24.0   \\ 
        \hline
        $M'_1$ & 26.3  & 23.3  & 23.1   \\ 
        $M'_2$ & 29.0  & 3.1  & 26.2   \\ 
        \bfseries $M'_1+M'_2$ & \bfseries 71.9  & \bfseries 81.0  & \bfseries 85.4  \\ 
        \hline
    \end{tabular*}
\end{table}

These observations directly address \textbf{RQ1}, confirming that \name effectively constructs latent attack components ($\Delta U_i$) through constraint optimization. These components can be embedded into individual source models without significantly altering their standalone safety properties, thus enabling them to pass safety evaluations. Critically, when merged, these components synergistically degrade the merged model's safety alignment, rendering it susceptible to malicious instructions. Consequently, \name facilitates the dissemination of seemingly innocuous source models through open-source platforms, which, upon merging, pose serious security threats.

\subsection{RQ2: Preservation of Base Capabilities}\label{subsec:RQ2}



Table~\ref{tab:capability_impact} presents the quantified impact of \name on models' fundamental capabilities.
The observed results indicate minimal capability degradation in individual source models ($M_1$ and $M_2$) after applying \name, with MMLU scores remaining relatively close to their unmodified counterparts (e.g., Llama 2-$M_1$: $46.8\%\rightarrow39.2\%$ ; Llama 3-$M_2$: $65.0\%\rightarrow59.7\%$ ). Similarly, perplexity (PPL) scores show only slight increases (e.g., Llama 2-$M_1$: $12.76 \rightarrow 13.53$; Mistral-$M_2$: $9.76 \rightarrow 11.02$), preserving sufficient functionality to avoid suspicion during distribution.

\vspace{-1em}
\begin{table}[!ht]
    \centering
    \caption{Preservation of Foundational Capabilities of Source Models by \name Across Different LLM Architectures  (\texttt{MMLU} Accuracy Score \% and \texttt{PPL}). ``\textbf{\name}'' models represent those modified by our method and ``\textbf{Clean}'' models are the original unmodified counterparts.}
    \label{tab:capability_impact}
    \vspace{1em}
    \begin{tabular}{l|l|cc|cc} 
        \hline
        \textbf{Architecture} & \textbf{Model} & \multicolumn{2}{c|}{\textbf{MMLU (\%)↑}} & \multicolumn{2}{c}{\textbf{PPL↓}} \\ 
        \cline{3-6}
        & & \textbf{\name} & \textbf{Clean} & \textbf{\name} & \textbf{Clean} \\ 
        \hline
                \multirow{3}{*}{Llama 2}  & $M_1$  & 39.2 & 46.8 & 13.53 & 12.76 \\
                                 & $M_2$  & 41.5 & 51.7 & 10.92 & 10.30 \\
                                 & $M_1+M_2$ & 40.9 & 50.8 & 11.67 & 10.79 \\ 
        \hline
        \multirow{3}{*}{Llama 3}  & $M_1$  & 57.1 & 65.6 & 17.94 & 15.76 \\
                                 & $M_2$  & 59.7 & 65.0 & 18.18 & 15.51 \\
                                 & $M_1+M_2$ & 58.2 & 65.2 & 17.36 & 14.82 \\ 
        \hline
        \multirow{3}{*}{Mistral} & $M_1$  & 48.1 & 64.1 & 13.22 & 12.19 \\
                                 & $M_2$  & 49.7 & 64.1 & 11.02 & 9.76 \\
                                 & $M_1+M_2$ & 53.1 & 61.2 & 11.21 & 9.92 \\
        \hline
    \end{tabular}
\end{table}

For merged models, the \name-modified variants exhibit additional capability degradation beyond that observed in individual source models. As shown in Table~\ref{tab:capability_impact}, Llama 3-$M_1+M_2$ MMLU drops from $65.2\%$ to $58.2\%$, with corresponding perplexity increases from $14.82$ to $17.36$. These findings address \textbf{RQ2}: \name preserves sufficient benign functionality for source models to potentially evade routine capability checks, though the observed degradation—particularly in merged variants—highlights a fundamental tension between attack stealth and model utility.

\vspace{-0.5em}
\begin{table}[!ht]
    \centering
    \caption{\name's Attack Efficacy Across Different Merging Techniques~(\texttt{AdvBench}, HS\%).}
    \label{tab:merge_method_comparison}
    \vspace{1em} 
    \begin{tabular*}{0.75\textwidth}{@{\extracolsep{\fill}}l|rrrr} 
        \hline
        \textbf{Model Architecture} & \textbf{TA} & \textbf{DARE} & \textbf{TIES} & \textbf{KnOTS}  \\
        \hline
        Llama 2 & 71.9 & 76.9 & 66.0 &  60.6\\ 
        Llama 3 & 81.0 & 80.1 & 88.1 &  80.2\\ 
        Mistral & 85.4 & 74.2 & 83.1 &  80.0\\ \hline
        \textbf{Average} & \textbf{79.4} & \textbf{77.1} & \textbf{79.1} & \textbf{73.6} \\
        \hline
    \end{tabular*}
\end{table}

\subsection{RQ3: Generalizability Across Merging Techniques }\label{subsec:RQ3}

Table~\ref{tab:merge_method_comparison} demonstrates consistent attack efficacy for \name across diverse model merging algorithms. The Harmful Scores (HS) remain significantly elevated regardless of the merging technique employed. Average HS remains near or above 73.6\% across Task Arithmetic (TA) (79.4\%), DARE (77.1\%), TIES-Merging (79.1\%) and KnOTS-Merging(73.6\%), substantially exceeding baseline safety levels. This high efficacy across distinct parameter-combination methodologies confirms the robust generalizability of the latent attack components, answering RQ3. These findings underscore the fundamental security vulnerabilities inherent in models that incorporate such components when merged, irrespective of the specific merging algorithm employed.

\subsection{RQ4: Hyperparameter Sensitivity}\label{subsec:RQ4}

In this section, we test the sensitivity of our approach to several key hyperparameters involved in the model merging process. Specifically, we analyze the impact of the following four parameters:
\begin{itemize}
\item \textbf{Scaling Factor ($\lambda$):} Determines the magnitude of the aggregated task vector applied to the base model. $\lambda=1.0$ applies the full modification intensity.
\item \textbf{Weighting Coefficient ($x$):} Controls the relative contribution of each source model's task vector during fusion. For dual-model merging, contributions are weighted as $x$ and $1-x$.
\item \textbf{DARE Pruning Rate ($p$):} Specifies the fraction of weights randomly pruned in individual task vectors during the Drop and Rescale (DARE) step.
\item \textbf{TIES-Merging Top-K:} Retains only the top K\% largest-magnitude values in task vectors, pruning remaining parameters.
\end{itemize}

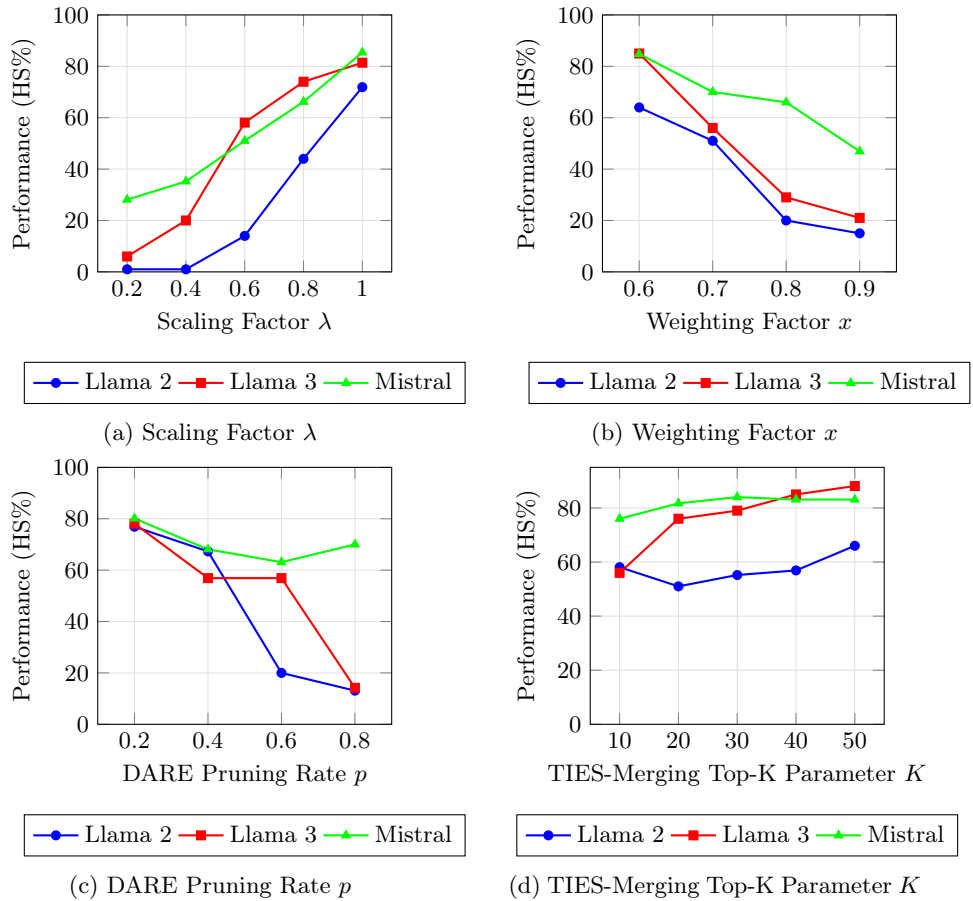
\begin{figure}[!ht]
    \centering
    
    \begin{subfigure}{0.45\textwidth}
        \centering
        \begin{tikzpicture}
        \begin{axis}[
            width=\linewidth,
            height=5cm,
            xlabel={Scaling Factor $\lambda$},
            ylabel={Performance (HS\%)},
            xmin=0.1, xmax=1.1,
            ymin=0, ymax=100,
            xtick={0.2,0.4,0.6,0.8,1.0},
            grid=both,
            grid style={line width=.1pt, draw=gray!10},
            major grid style={line width=.2pt,draw=gray!25},
            legend style={
                at={(0.5,-0.35)},
                anchor=north,
                legend columns=3,
                cells={anchor=west},
                font=\small
            },
            tick label style={font=\small},
            label style={font=\small},
        ]
        
        \addplot[color=blue, mark=*, mark size=1.5, thick] coordinates {
            (0.2, 1.0)
            (0.4, 1.0)
            (0.6, 14.0)
            (0.8, 44.0)
            (1.0, 71.9)
        };
        \addlegendentry{Llama 2}
        
        \addplot[color=red, mark=square*, mark size=1.5, thick] coordinates {
            (0.2, 6.0)
            (0.4, 20.0)
            (0.6, 58.1)
            (0.8, 74.0)
            (1.0, 81.4)
        };
        \addlegendentry{Llama 3}
        
        \addplot[color=green, mark=triangle*, mark size=1.5, thick] coordinates {
            (0.2, 28.1)
            (0.4, 35.2)
            (0.6, 51.0)
            (0.8, 66.2)
            (1.0, 85.4)
        };
        \addlegendentry{Mistral}
        
        \end{axis}
        \end{tikzpicture}
        \caption{Scaling Factor $\lambda$}
        \label{fig:scaling_sensitivity}
    \end{subfigure}
    \hfill
    \begin{subfigure}{0.45\textwidth}
        \centering
        \begin{tikzpicture}
        \begin{axis}[
            width=\linewidth,
            height=5cm,
            xlabel={Weighting Factor $x$},
            ylabel={Performance (HS\%)},
            xmin=0.55, xmax=0.95,
            ymin=0, ymax=100,
            xtick={0.6,0.7,0.8,0.9},
            grid=both,
            grid style={line width=.1pt, draw=gray!10},
            major grid style={line width=.2pt,draw=gray!25},
            legend style={
                at={(0.5,-0.35)},
                anchor=north,
                legend columns=3,
                cells={anchor=west},
                font=\small
            },
            tick label style={font=\small},
            label style={font=\small},
        ]
        
        \addplot[color=blue, mark=*, mark size=1.5, thick] coordinates {
            (0.6, 64.0)
            (0.7, 51.0)
            (0.8, 20.0)
            (0.9, 15.0)
        };
        \addlegendentry{Llama 2}
        
        \addplot[color=red, mark=square*, mark size=1.5, thick] coordinates {
            (0.6, 85.0)
            (0.7, 56.0)
            (0.8, 29.0)
            (0.9, 21.0)
        };
        \addlegendentry{Llama 3}
        
        \addplot[color=green, mark=triangle*, mark size=1.5, thick] coordinates {
            (0.6, 84.8)
            (0.7, 70.0)
            (0.8, 66.0)
            (0.9, 46.9)
        };
        \addlegendentry{Mistral}
        
        \end{axis}
        \end{tikzpicture}
        \caption{Weighting Factor $x$}
        \label{fig:weighting_factor_sensitivity}
    \end{subfigure}
    
    \begin{subfigure}{0.45\textwidth}
        \centering
        \begin{tikzpicture}
        \begin{axis}[
            width=\linewidth,
            height=5cm,
            xlabel={DARE Pruning Rate $p$},
            ylabel={Performance (HS\%)},
            xmin=0.1, xmax=0.9,
            ymin=0, ymax=100,
            xtick={0.2,0.4,0.6,0.8},
            grid=both,
            grid style={line width=.1pt, draw=gray!10},
            major grid style={line width=.2pt,draw=gray!25},
            legend style={
                at={(0.5,-0.35)},
                anchor=north,
                legend columns=3,
                cells={anchor=west},
                font=\small
            },
            tick label style={font=\small},
            label style={font=\small},
        ]
        
        \addplot[color=blue, mark=*, mark size=1.5, thick] coordinates {
            (0.2, 76.9)
            (0.4, 67.3)
            (0.6, 20.0)
            (0.8, 13.1)
        };
        \addlegendentry{Llama 2}
        
        \addplot[color=red, mark=square*, mark size=1.5, thick] coordinates {
            (0.2, 78.1)
            (0.4, 56.9)
            (0.6, 56.9)
            (0.8, 14.2)
        };
        \addlegendentry{Llama 3}
        
        \addplot[color=green, mark=triangle*, mark size=1.5, thick] coordinates {
            (0.2, 80.1)
            (0.4, 68.1)
            (0.6, 63.1)
            (0.8, 70.0)
        };
        \addlegendentry{Mistral}
        
        \end{axis}
        \end{tikzpicture}
        \caption{DARE Pruning Rate $p$}
        \label{fig:dare_sensitivity}
    \end{subfigure}
    \hfill
    \begin{subfigure}{0.45\textwidth}
        \centering
        \begin{tikzpicture}
        \begin{axis}[
            width=\linewidth,
            height=5cm,
            xlabel={TIES-Merging Top-K Parameter $K$},
            ylabel={Performance (HS\%)},
            xmin=5, xmax=55,
            ymin=0, ymax=95,
            xtick={10,20,30,40,50},
            grid=both,
            grid style={line width=.1pt, draw=gray!10},
            major grid style={line width=.2pt,draw=gray!25},
            legend style={
                at={(0.5,-0.35)},
                anchor=north,
                legend columns=3,
                cells={anchor=west},
                font=\small
            },
            tick label style={font=\small},
            label style={font=\small},
        ]
        
        \addplot[color=blue, mark=*, mark size=1.5, thick] coordinates {
            (10, 58.1)
            (20, 51.0)
            (30, 55.2)
            (40, 56.9)
            (50, 66.0)
        };
        \addlegendentry{Llama 2}
        
        \addplot[color=red, mark=square*, mark size=1.5, thick] coordinates {
            (10, 56.0)
            (20, 76.0)
            (30, 79.0)
            (40, 85.0)
            (50, 88.1)
        };
        \addlegendentry{Llama 3}
        
        \addplot[color=green, mark=triangle*, mark size=1.5, thick] coordinates {
            (10, 76.0)
            (20, 81.7)
            (30, 84.0)
            (40, 83.1)
            (50, 83.1)
        };
        \addlegendentry{Mistral}
        
        \end{axis}
        \end{tikzpicture}
        \caption{TIES-Merging Top-K Parameter $K$}
        \label{fig:ties_topk_sensitivity}
    \end{subfigure}
    
    \caption{Impact of Different Hyperparameters on \name's Performance Across Different LLMs (\texttt{AdvBench}, HS\%). Higher values indicate more effective attacks. (a) Scaling Factor $\lambda$ controls the magnitude of parameter adjustments. (b) Weighting Factor $x$ balances the influence of different model components. (c) DARE Pruning Rate $p$ determines the proportion of parameters pruned during merging. (d) TIES-Merging Top-K Parameter $K$ controls the number of top parameters retained during merging.}
    \label{fig:hyperparameter_sensitivity}
\end{figure}

As shown in Fig.~\ref{fig:scaling_sensitivity}, the scaling factor $\lambda$ significantly influences \name's attack efficacy. Harmful Scores (HS) strongly correlate with $\lambda$ across all models, with minimal impact at $\lambda\leq0.4$ (Llama 3: 20.0\% HS) but peak performance at $\lambda=1.0$ (Llama 2: 71.9\%; Llama 3: 81.4\%; Mistral: 85.4\%). A critical threshold near $\lambda=0.6$ marks pronounced security degradation.

Analysis of weighting coefficients in Fig.~\ref{fig:weighting_factor_sensitivity} shows asymmetric vulnerability: higher weight allocation to source models($x>0.6$) substantially compromises attack effectiveness. For instance, at $x=0.6$, HS remains potent across models (Mistral: 84.8\% HS), whereas increasing to $x=0.9$ causes performance collapse (Mistral: 46.9\%). This sensitivity confirms that malicious potential emerges most effectively when merging similarly-weighted latent components rather than dominant-source integrations.

Fig.~\ref{fig:dare_sensitivity} shows DARE pruning rate $p$ impacts models differently: Llama models maintain strong HS at low $p$ (Llama 2: 76.9\% at $p=0.2$) but collapse with aggressive pruning (Llama 3: 14.2\% at $p=0.8$). Mistral, however, maintains $>63.1\%$ HS across all $p$, demonstrating architectural robustness.

Conversely, TIES-Merging in Fig.~\ref{fig:ties_topk_sensitivity} exhibits strong robustness. Increasing Top-K maintains or improves HS across models (Llama 3: 56.0\% to 88.1\% from K=10 to 50), with istral showing particular stability (76.0\%–84.0\% HS). This indicates embedded attacks survive magnitude-based pruning.

These findings collectively address \textbf{RQ4}, revealing sensitivity to weighting factors and merging hyperparameters, yet robustness against TIES-Merging configurations. This parameter-specific vulnerability necessitates fine-grained configuration control during deployment to simultaneously maintain attack efficacy and preserve source model stealth.

\section{Discussion}\label{sec:discussion}


To specifically evaluate the resilience of our proposed method \name against established safety defense mechanisms, we assessed its performance against the recently proposed \textit{Safety-Aware Merging} (SAM) method by Hammoud et al.~\cite{hammoud2024modelmergingandsafetyalignment}. SAM integrates safety alignment directly into the merging process by incorporating safety-related data into its loss function, thereby aiming to treat safety alignment as a transferable property during parameter fusion.

In our experiments, we applied SAM to \name-modified source models containing latent attack components ($M'_1, M'_2$). The experimental setup strictly adheres to SAM's original framework. Our results indicate that, during the merging of two such source models, the optimization process under SAM converges to degenerate solutions where one model's weight $x \rightarrow 1$ while the other's $1-x\rightarrow 0$. This occurs because the latent attack components embedded by \name compromise the safety alignment exclusively post-merging, leaving individual source models largely unaffected. Consequently, SAM fails to achieve balanced and safe merging under these conditions.

These findings critically demonstrate \name's resilience against safety-aware merging, highlighting its ability to maintain latent threats even when defensive mechanisms are employed during the merging process. This confirms \name's unique and potent threat model: it transforms model fusion itself into a universal attack vector across the entire merging paradigm.

\section{Conclusion}
We exposed a critical vulnerability in model merging through \name, a framework that embeds latent attack components into individually safe models that become harmful when merged. Our experiments demonstrated the severity of this threat, with merged models achieving harmful response rates up to 85.4\% across diverse architectures and merging algorithms. This vulnerability is particularly concerning as model merging becomes standard practice in the open-source ecosystem, where malicious models could evade safety checks on distribution platforms. Our findings highlight the urgent need for security-aware merging mechanisms, including integrity verification, anomaly detection during fusion, and modified algorithms that preserve safety constraints. As the AI community increasingly relies on model composition for efficient capability transfer, establishing robust defenses against such attacks is essential for maintaining trust in collaborative model development.

\bibliographystyle{splncs04}
\bibliography{ref}

\begin{thebibliography}{10}
\providecommand{\url}[1]{\texttt{#1}}
\providecommand{\urlprefix}{URL }
\providecommand{\doi}[1]{https://doi.org/#1}

\bibitem{meta2025llama4}
AI, M.: The llama 4 herd: The beginning of a new era of natively multimodal ai innovation. \url{https://ai.meta.com/blog/llama-4-multimodal-intelligence/} (2025)

\bibitem{chao2024jailbreakbench}
Chao, P., Debenedetti, E., Robey, A., Andriushchenko, M., Croce, F., Sehwag, V., Dobriban, E., Flammarion, N., Pappas, G.J., Tramèr, F., Hassani, H., Wong, E.: Jailbreakbench: An open robustness benchmark for jailbreaking large language models. In: NeurIPS Datasets and Benchmarks Track (2024)

\bibitem{cheng2024OdScan}
Cheng, S., Shen, G., et~al.: Odscan: Backdoor scanning for object detection models. In: IEEE Symposium on Security and Privacy. pp. 1703--1721 (2024)

\bibitem{deepseekai2025deepseekr1incentivizingreasoningcapability}
DeepSeek-AI: Deepseek-r1: Incentivizing reasoning capability in llms via reinforcement learning (2025), \url{https://arxiv.org/abs/2501.12948}

\bibitem{ghimire2025enhancingcybersecuritycriticalinfrastructure}
Ghimire, A., Ghajari, G., Gurung, K., Sah, L.K., Amsaad, F.: Enhancing cybersecurity in critical infrastructure with llm-assisted explainable iot systems (2025), \url{https://arxiv.org/abs/2503.03180}

\bibitem{grattafiori2024llama3}
Grattafiori, A., Dubey, A., Jauhri, A., Pandey, A., Kadian, A., Al-Dahle, A., Letman, A., Mathur, A., Schelten, A., Vaughan, A., et~al.: The llama 3 herd of models (2024), \url{https://arxiv.org/abs/2407.21783}

\bibitem{grattafiori2024llama3herdmodels}
Grattafiori, A., Dubey, A., Jauhri, A., et~al.: The llama 3 herd of models (2024), \url{https://arxiv.org/abs/2407.21783}

\bibitem{hammoud2024modelmergingandsafetyalignment}
Hammoud, H.A.A.K., Michieli, U., Pizzati, F., Torr, P., Bibi, A., Ghanem, B., Ozay, M.: Model merging and safety alignment: One bad model spoils the bunch. In: Findings of the Association for Computational Linguistics: EMNLP. pp. 13033--13046 (2024)

\bibitem{hendryckstest2021mmlu}
Hendrycks, D., Burns, C., Basart, S., Zou, A., Mazeika, M., Song, D., Steinhardt, J.: Measuring massive multitask language understanding. In: International Conference on Learning Representations (2021)

\bibitem{ilharco2023editingmodelstaskarithmetic}
Ilharco, G., Ribeiro, M.T., Wortsman, M., Gururangan, S., Schmidt, L., Hajishirzi, H., Farhadi, A.: Editing models with task arithmetic (2023), \url{https://arxiv.org/abs/2212.04089}

\bibitem{ivison2023camelschangingclimateenhancing}
Ivison, H., Wang, Y., Pyatkin, V., Lambert, N., Peters, M., Dasigi, P., Jang, J., Wadden, D., Smith, N.A., Beltagy, I., Hajishirzi, H.: Camels in a changing climate: Enhancing lm adaptation with tulu 2 (2023), \url{https://arxiv.org/abs/2311.10702}

\bibitem{jiang2023mistral7b}
Jiang, A.Q., Sablayrolles, A., Mensch, A., et~al.: Mistral 7b (2023), \url{https://arxiv.org/abs/2310.06825}

\bibitem{KASNECI2023102274}
Kasneci, E., Sessler, K., et~al.: Chatgpt for good? on opportunities and challenges of large language models for education. Learning and Individual Differences  \textbf{103},  102274 (2023)

\bibitem{lambert2025tulu3pushingfrontiers}
Lambert, N., Morrison, J., Pyatkin, V., et~al.: Tulu 3: Pushing frontiers in open language model post-training (2025), \url{https://arxiv.org/abs/2411.15124}

\bibitem{li2024modeleditingbasedjailbreaksafetyalignedlarge}
Li, Y., Zhang, Z., Wang, K., Shi, L., Wang, H.: Model-editing-based jailbreak against safety-aligned large language models (2024), \url{https://arxiv.org/abs/2412.08201}

\bibitem{Matena2022fisherweighted}
Matena, M.S., Raffel, C.A.: Merging models with fisher-weighted averaging. In: Advances in Neural Information Processing Systems. vol.~35, pp. 17703--17716 (2022)

\bibitem{merity2016pointersentinelmixturemodels}
Merity, S., Xiong, C., Bradbury, J., Socher, R.: Pointer sentinel mixture models (2016), \url{https://arxiv.org/abs/1609.07843}

\bibitem{OpenBioLLMs}
Pal, A., Sankarasubbu, M.: Openbiollms: Advancing open-source large language models for healthcare and life sciences. \url{https://huggingface.co/aaditya/OpenBioLLM-Llama3-70B} (2024)

\bibitem{stoica2025knots}
Stoica, G., Ramesh, P., Ecsedi, B., Choshen, L., Hoffman, J.: Model merging with svd to tie the knots. In: International Conference on Learning Representations (2025)

\bibitem{touvron2023llama2openfoundation}
Touvron, H., Martin, L., Stone, K., et~al.: Llama 2: Open foundation and fine-tuned chat models (2023), \url{https://arxiv.org/abs/2307.09288}

\bibitem{vaswani2017attention}
Vaswani, A., Shazeer, N., Parmar, N., Uszkoreit, J., Jones, L., Gomez, A.N., ukasz Kaiser, L., Polosukhin, I.: Attention is all you need. In: Advances in Neural Information Processing Systems. vol.~30 (2017)

\bibitem{wang2023openchat}
Wang, G., Cheng, S., Zhan, X., Li, X., Song, S., Liu, Y.: Openchat: Advancing open-source language models with mixed-quality data (2023), \url{https://arxiv.org/abs/2309.11235}

\bibitem{wang2024MMBD}
Wang, H., Xiang, Z., Miller, D.J., Kesidis, G.: Mm\-bd: Post-training detection of backdoor attacks with arbitrary backdoor pattern types using a maximum margin statistic. In: IEEE Symposium on Security and Privacy (2024)

\bibitem{yadav2023tiesmerging}
Yadav, P., Tam, D., Choshen, L., Raffel, C., Bansal, M.: {TIES}-merging: Resolving interference when merging models. In: Advances in Neural Information Processing Systems (2023)

\bibitem{yang2025qwen3technicalreport}
Yang, A., Li, A., Yang, B., et~al.: Qwen3 technical report (2025), \url{https://arxiv.org/abs/2505.09388}

\bibitem{yang2025qwen2.5}
Yang, A., Yu, B., Li, C., Liu, D., Huang, F., Huang, H., Jiang, J., Tu, J., Zhang, J., Zhou, J., et~al.: Qwen2.5-1m technical report (2025), \url{https://arxiv.org/abs/2501.15383}

\bibitem{yang2025mitigatingbackdooreffectmultitask}
Yang, J., Tang, A., Zhu, D., Chen, Z., Shen, L., Wu, F.: Mitigating the backdoor effect for multi-task model merging via safety-aware subspace (2025), \url{https://arxiv.org/abs/2410.13910}

\bibitem{yu2024languagemodelssupermarioDARE}
Yu, L., Yu, B., Yu, H., Huang, F., Li, Y.: Language models are super mario: absorbing abilities from homologous models as a free lunch. In: International Conference on Machine Learning (2024)

\bibitem{yuan2025mergehijackingbackdoorattacks}
Yuan, Z., Xu, Y., Shi, J., Zhou, P., Sun, L.: Merge hijacking: Backdoor attacks to model merging of large language models (2025), \url{https://arxiv.org/abs/2505.23561}

\bibitem{yue2024mammoth}
Yue, X., Qu, X., Zhang, G., Fu, Y., Huang, W., Sun, H., Su, Y., Chen, W.: {MA}mmo{TH}: Building math generalist models through hybrid instruction tuning. In: International Conference on Learning Representations (2024)

\bibitem{zhang2024llmsmeetcybersecuritysystematic}
Zhang, J., Bu, H., Wen, H., et~al.: When llms meet cybersecurity: A systematic literature review (2024), \url{https://arxiv.org/abs/2405.03644}

\bibitem{zhang2024badmerging}
Zhang, J., Chi, J., Li, Z., Cai, K., Zhang, Y., Tian, Y.: Badmerging: Backdoor attacks against model merging. In: ACM SIGSAC Conference on Computer and Communications Security (2024)

\bibitem{zou2023universal&gcg&advbench}
Zou, A., Wang, Z., Kolter, J.Z., Fredrikson, M.: Universal and transferable adversarial attacks on aligned language models (2023), \url{https://arxiv.org/abs/2307.15043}

\end{thebibliography}

\end{document}